\documentstyle[aps,pra,amssymb,twocolumn,epsfig,floats]{revtex}

\def\Red  {}
\def\Black{}
\def\Blue {}
\begin{document}
\draft

\title{\Red The interference of a nonclassical light pulse with a coherent one\\
and the sub-Poissonian statistics formation.\Black}
\author{F. Popescu${}^{\star}$}
\address{
        \mbox{~}\\ [-1mm]
        {\sf E-mail: ${}^{\star}$ florentin\_p@hotmail.com}}
\date{June 26, 2000}
\twocolumn[\widetext
\begin{@twocolumnfalse}
\maketitle
\begin{abstract}\Blue
Using the theoretical model of the optical beam-splitter, the interference of the
self-phase modulated ultrashort light pulse (SPM-USP) with the coherent one is
investigated. It is found that, the choice of the coefficient of transmission of the
beam-splitter allows one to get the spectra of quadrature fluctuations with forms of
interest to us. It is shown that, the choice of the geometrical phase gives one the
control of the position of the ellipse of squeezing in the quadrature space $XY$. The
extended Mandel parameter is introduced and the photon statistics is scanned at all
frequencies. It is established that, the sub- and super-Poissonian statistics
formation can be determined by the choice of the nonlinear phase addition and initial
linear phase shift between pulses. It is also shown that, the self-phase modulation
(SPM) leads to the additional modulation of total photon number at the outputs of the
beam-splitter.\\ \Black \pacs{PACS numbers: 42.50.A, 42-50.Dv, 42-25.H}
\end{abstract}
\vspace{0.5cm} \narrowtext
\end{@twocolumnfalse}]

\section{Introduction}
The formation of {\em ultrashort light pulses} (USPs) with suppressed  photon fluctuations
remains in the focus of considerable attention. The formation and application
of USPs in a nonclassical state make possible to combine in experiments a high
time resolution with a low level of fluctuations. There are some methods to obtain
the USPs in the nonclassical state.

Parametric amplification is a technique that is most extensively used for the
production of the USPs in the nonclassical state. In the case of degenerate
three-frequency parametric amplification, quadrature-squeezed light is produced.
However, this light is found to have super-Poissonian photon statistics directly at
the output of an amplifier, and needs interferometers to transform it to one with
sub-Poissonian statistics.

One of the most interesting methods for production of USPs in the nonclassical state
is the {\em self-phase modulation} (SPM) in a nonlinear inertial medium
\cite{POP00a,POP00b,POP99}. In this case the USP in the quadrature-squeezing state
can be formed with conservation of the photon statistics \cite{Kitagava}. The SPM
itself is not accompanied by a change in photon statistics. With the aid of nonlinear
optical devices in the presence of the SPM one can obtain light with sub-Poissonian
photon statistics (see \cite{Ahmanov}).

For the first time in \cite{POP00c} a simple method for the production of the USPs
with suppressed photon number fluctuations was considered, which is based on the SPM
of a USP in a nonlinear inertial medium and subsequent transmission of a pulse
through a dispersive optical element. The accurate calculation of this process
was conducted making use of  technique developed in \cite{POP00a,POP00b,POP99}.

In the present paper, the interference of the SPM-USP with the coherent USP is
analysed. The interference is realized using the optical beam-splitter and the
process under consideration is analysed in the framework of the quantum theory of
SPM of USPs developed in \cite{POP00a,POP00b,POP99}.

The aim of the present article is to analyse the spectra of quantum fluctuations of
quadrature components and the spectra of quantum fluctuations of the photon number of
the pulses at the outputs of an optical beam-splitter. An extended Mandel parameter
will be introduced in order to analyse the photon statistics at all frequencies. The
modulation of the total photon number will be analysed too.

\section{Spectra of quantum  fluctuations of quadratures of SPM-USP{\scriptsize{s}}}
For the first time in \cite{POP00a,POP00b,POP99} the consequently quantum theory of
self-action of USPs in nonlinear inertial medium based on the algebra of
time-dependent Bose-operators has been developed. When analysed from the quantum
point of view, SPM of USPs is described by the expression \cite{POP00a,POP00b,POP99}
\begin{equation}\label{operator}
\hat{A}(t,z)=e^{\hat{O}(t)}\,\hat{A}_{0}(t),
\end{equation}
where $\hat{A}(t,z)$ is the annihilation Bose-operator in the given cross section $z$
of the nonlinear inertial medium, $\hat{O}(t)=i\gamma q[\hat{n}_{0}(t)]$ and $\gamma
q[\hat{n}_{0}(t)]$ is nonlinear phase incursion. The permutation relation takes place:
\begin{equation}\label{algebra}
\hat{A}_{0}(t_{1})e^{\hat{O}(t_{2})}=e^{\hat{O}(t_{2})+{G}(t_{2}-t_{1})}\hat{A}_{0}(t_{1}),
\end{equation}
and the theorem of the normal ordering is valid
\begin{equation}\label{teor}
e^{\hat{O}(t)}=\hat{\mathbf{N}}\exp\left\{\int_{-\infty}^{\infty}
\left[e^{{\mathcal{H}}(\theta)}-1\right]\hat{n}_{0}(t-\theta\tau_{r})d\theta\right\},
\end{equation}
where ${G}(t_2-t_1)=i\gamma h(t_2-t_1)$, $h(t)=H(|t|)$,
${\mathcal{H}}(\theta)=i\gamma \widetilde{h}(\theta)$,
$\widetilde{h}(\theta)=\tau_{r}h(\theta\tau_r)$, $\tau_{r}$ is the relaxation time of
the nonlinear medium and $\hat{\mathbf{N}}$ is the operator of normal ordering. Here
$H(t)$ represents the nonlinear response function of the nonlinear medium ($H(t)>0$
for $t>0$ and $H(t)=0$ for $t<0$). Eqs.(\ref{operator}-\ref{teor}) are written in the
moving frame $t=t'-z/u$, $z=z'$, where $t'$ is the time measured in co-moving frame
and $u$ is the speed of a pulse in the nonlinear inertial medium.

The quadrature components of the light pulse are defined by the expressions
\begin{mathletters}
\begin{eqnarray}
\hat{X}(t,z)&=&[\hat{A}(t,z)+\hat{A}^{+}(t,z)]/2,\label{ceas1}\\
\hat{Y}(t,z)&=&[\hat{A}(t,z)-\hat{A}^{+}(t,z)]/2i,\label{ceas2}
\end{eqnarray}
\end{mathletters}
and for average values of quadratures we have \cite{POP00a,POP00b,POP99}
\begin{mathletters}
\begin{eqnarray}
\langle\hat{X}(t,z)\rangle&=&|\alpha_{0}(t)|e^{-\mu(t)}\cos{\Phi(t)},\label{xip}\\
\langle\hat{Y}(t,z)\rangle&=&|\alpha_{0}(t)|e^{-\mu(t)}\sin{\Phi(t)}.\label{xix}
\end{eqnarray}
\end{mathletters}
In (\ref{xip}-\ref{xix}) we denoted $\Phi(t)=\psi(t)+\varphi(t)$,
$\psi(t)=2\gamma|\alpha_{0}(t)|^{2}=2\gamma\bar{n}_{0}(t)$ and
$\mu(t)=\gamma^{2}\bar{n}_{0}(t)=\gamma\psi(t)/2$. The correlation functions of the
quadrature components must be introduced as \cite{POP00a,POP00b,POP99}
\begin{mathletters}
\begin{eqnarray}
R_{X}(t,t\smash{+}\tau)\!=\!\frac{1}{2}\!\!&\Bigl[&\!\!\!
\langle\hat{X}(t,z)\hat{X}(t\smash{+}\tau,z)\rangle\smash{+}
\langle\hat{X}(t\smash{+}\tau,z)\hat{X}(t,z)\rangle\nonumber \\
&-&2\langle\hat{X}(t,z)\rangle\langle\hat{X}(t\smash{+}\tau,z)\rangle\Bigl],\label{king1}\\
R_{Y}(t,t\smash{+}\tau)\!=\!\frac{1}{2}\!\!&\Bigl[&\!\!\!
\langle\hat{Y}(t,z)\hat{Y}(t\smash{+}\tau,z)\rangle\smash{+}\langle
\hat{Y}(t\smash{+}\tau,z)\hat{Y}(t,z)\rangle\nonumber \\
&-&2\langle\hat{Y}(t,z)\rangle\langle\hat{Y}(t\smash{+}\tau,z)\rangle\Bigl],\label{king2}
\end{eqnarray}
\end{mathletters}
where $\tau=t_{1}\smash{-}t$ ($t_{1}$ is an arbitrary moment of time, $t_{1}>t$).
Using algebra of time-dependent Bose-operators (see (\ref{algebra}-\ref{teor})) for
the correlation functions we find
\begin{mathletters}
\begin{eqnarray}
R_{X}(t,t+\tau)=\frac{1}{4}\Bigl\{\delta(\tau)&-&\psi(t)h(\tau)\sin{2\Phi(t)}\nonumber\\
&+&\psi^{2}(t)g(\tau)\sin^{2}{\Phi(t)}\Bigl\},\label{rer}\\
R_{Y}(t,t+\tau)=\frac{1}{4}\Bigl\{\delta(\tau)&+&\psi(t)h(\tau)\sin{2\Phi(t)}\nonumber\\
&+&\psi^{2}(t)g(\tau)\cos^{2}{\Phi(t)}\Bigl\},\label{yer}
\end{eqnarray}
\end{mathletters}
where we denoted $h(\tau)\smash{=}\tau^{-1}_{r}\exp{\{-|\tau|/\tau_{r}\}}$,
$g(\tau)=\tau^{-1}_{r}(1+|\tau|/\tau_{r})\exp{\{-|\tau|/\tau_{r}\}}$ and considered
that nonlinear inertial medium is of a Kerr type. To get the expressions
(\ref{rer}-\ref{yer}), the approximations $\tau_r\ll\tau_p$ , $\gamma\ll1$ have been
used ($\tau_{p}$ is the pulse duration). The spectra of the fluctuations of
quadrature components are
\begin{mathletters}
\begin{eqnarray}
S_{X}(\Omega,t)=\frac{1}{4}\bigl[1&-&2\psi(t)L(\Omega)\sin{2\Phi(t)}\nonumber\\
&{}&+4\psi^{2}(t)L^{2}(\Omega)\sin^{2}{\Phi(t)}\bigl],\label{sio}\\
S_{Y}(\Omega,t)=\frac{1}{4}\bigl[1&+&2\psi(t)L(\Omega)\sin{2\Phi(t)}\nonumber\\
&{}&+4\psi^{2}(t)L^{2}(\Omega)\cos^{2}{\Phi(t)}\bigl],\label{ert}
\end{eqnarray}
\end{mathletters}
where we denoted $\Omega=\omega\tau_{r}$ and $L(\Omega)=1/[1+\Omega^{2}]$. At initial
phase chosen optimal at the frequency $\Omega_{0}$ (see \cite{POP00a,POP00b,POP99})
\begin{equation}
\varphi_{0}(t)=\frac{1}{2}\arctan{\left\{\frac{1}{\psi(t)L(\Omega_{0})}\right\}}-
\psi(t)\label{phase}
\end{equation}
the spectra (\ref{sio}-\ref{ert}) have the forms
\begin{mathletters}
\begin{eqnarray}
S_{X}(\Omega_{0},t)&=&\frac{1}{4}\bigl[\sqrt{1+\psi^{2}(t)L^{2}(\Omega_{0})}
-\psi(t)L(\Omega_{0})\bigl]^{2},\label{So}\\
S_{Y}(\Omega_{0},t)&=&\frac{1}{4}\bigl[\sqrt{1+\psi^{2}(t)L^{2}(\Omega_{0})}
+\psi(t)L(\Omega_{0})\bigl]^{2}.\label{Soe}
\end{eqnarray}
\end{mathletters}
At any frequency $\Omega$ the spectra (\ref{sio}-\ref{ert}) are given by:
\begin{mathletters}
\begin{eqnarray}
S_{X}(\Omega,t)&=&S_{X}(\Omega_{0},t)+\frac{1}{2}\psi(t)[L(\Omega)-L(\Omega_{0})]\nonumber\\
&{}&\times\Bigl\{[L(\Omega)\smash{+}L(\Omega_{0})]\psi(t)\smash{-}[1\smash{+}(L(\Omega)\smash{+}
L(\Omega_{0}))\nonumber\\ &{}&~\times
L(\Omega_{0})\psi^{2}(t)][1\smash{+}\psi^{2}(t)L^{2}(\Omega)]^{-1/2}\Bigl\},\label{sist}\\
S_{Y}(\Omega,t)&=&S_{Y}(\Omega_{0},t)+\frac{1}{2}\psi(t)[L(\Omega)-L(\Omega_{0})]\nonumber\\
&{}&\times\Bigl\{[L(\Omega)\smash{+}L(\Omega_{0})]\psi(t)\smash{+}[1\smash{+}(L(\Omega)\smash{+}
L(\Omega_{0}))\nonumber\\ &{}&~\times
L(\Omega_{0})\psi^{2}(t)][1\smash{+}\psi^{2}(t)L^{2}(\Omega)]^{-1/2}\Bigl\}.\label{sios}
\end{eqnarray}
\end{mathletters}

\begin{figure}[t]
   \begin{center}
       \leavevmode
       \epsfxsize=.48\textwidth
       \epsfysize=.4\textwidth
       \epsffile{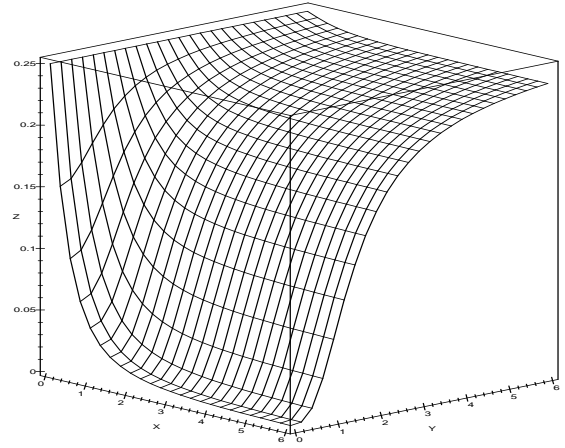}
   \end{center}
   \caption{Dispersion [Z] of the $X$-quadrature component of a pulse at time $t\smash{=}0$
as a function of the maximum nonlinear phase $\psi(0)$ [X] and the reduced frequency
$\Omega\smash{=}\omega\tau_r$ [Y], at values of the phase of the pulse which are
optimal for $\Omega_{0}\smash{=}0$.\label{fig1}}
\end{figure}

The spectra of the quantum fluctuations of squeezed $X$-quadrature at the initial
phase chosen optimal at the reduced frequency $\Omega_{0}=0$ are displayed in
Fig.\ref{fig1}.
{}From Fig.\ref{fig1} one can see that the squeezing of quantum fluctuations is
greatest at the frequency for which the phase of the initial pulse was chosen
optimal. In addition, in \cite{POP00a,POP00b,POP99} have been shown that the spectra
of quantum fluctuations of quadrature components can be controlled by the choice of
the phase of the initial light pulse and that, at the initial phase chosen optimal at
the frequency $\Omega_{0}=1$, the squeezing of quadrature component $X$ is maximum at
frequencies $\omega\approx1/\tau_{r}$ for $\psi(t)>1$.

The choice of the optimal phase as (\ref{phase}) means that in the quadrature's space
$XY$ the big axis of the eclipse of squeezing is parallel with the $Y$ axis and
consequently,  the squeezing of the $X$ quadrature is maximum (see (\ref{sist})). The
transformation of the uncertainty region of quadrature fluctuations, from the circle
for the initial coherent light pulse to the ellipse for the SPM-USP, always take
place. The choice of the phase as (\ref{phase}) means the orientation of the system
of quadrature squeezing observation $XY$ as the quantum fluctuations of $X$
quadrature are maximum suppressed.

\section{The interference of SPM-USP with a coherent one}
\begin{figure}[t]
   \begin{center}
       \leavevmode
       \epsfxsize=.5\textwidth
       \epsfysize=.5 \textwidth
       \epsffile{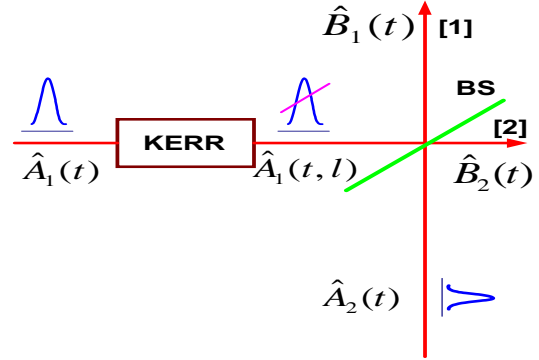}
   \end{center}
   \caption{The scheme of the optical beam-splitter (BS) in which the interference between the
nonclassical SPM-USP and coherent SPM is realized. [1] and [2] represent the BS
outputs. \label{bs}}
\end{figure}

In what follows we are interested in analysing of the interference between the
SPM-USP and coherent one. As a theoretical model we use the model of the symmetric
beam-splitter \cite{Ou} (double reflection angle of the incident SPM-USP is $\pi/2$).
The interference scheme is presented in Fig.\ref{bs}.  The input-output operator
transformations \cite{Ou} at the output \# $1$ are given by
\begin{mathletters}
\begin{eqnarray}
\hat{B}_{1}(t)&=&i\sqrt{R}\,\hat{A}_{1}(t,l)+\sqrt{T}\,\hat{A}_{2}(t),\label{transf1}\\
\hat{B}^{+}_{1}(t)&=&-i\sqrt{R}\,\hat{A}^{+}_{1}(t,l)+\sqrt{T}\,\hat{A}_{2}(t),\label{transf2}
\end{eqnarray}
\end{mathletters}
and at the output \# 2 they are
\begin{mathletters}
\begin{eqnarray}
\hat{B}_{2}(t)&=&\sqrt{T}\,\hat{A}_{1}(t,l)+i\sqrt{R}\,\hat{A}_{2}(t),\label{transf1a}\\
\hat{B}^{+}_{2}(t)&=&\sqrt{T}\,\hat{A}^{+}_{1}(t,l)-i\sqrt{R}\,\hat{A}^{+}_{2}(t),\label{transf2a}
\end{eqnarray}
\end{mathletters}
where $l$ is the length of the nonlinear inertial medium and $R$ is the coefficient
of reflection, $R+T=1$.
\subsection{The spectra of quantum fluctuations of quadratures at the beam-splitter output \# $1$}
We define the quadrature components at the output \# $1$ of the beam splitter as (see
(\ref{ceas1}-\ref{ceas2}))
\begin{mathletters}
\begin{eqnarray}
\hat{\mathcal{X}}_{1}(t)&=&[\hat{B}_{1}(t)+\hat{B}^{+}_{1}(t)]/2,\label{pow1}\\
\hat{\mathcal{Y}}_{1}(t)&=&[\hat{B}_{1}(t)-\hat{B}^{+}_{1}(t)]/2i.\label{pow2}
\end{eqnarray}
\end{mathletters}
For average values of the quadrature components (\ref{pow1}-\ref{pow2}) we find
\begin{mathletters}
\begin{eqnarray}
\langle\hat{\mathcal{X}}_{1}(t)\rangle&\smash{=}&|\alpha_{0}(t)|e^{-\mu(t)}
[\sqrt{T}\cos\varphi_{2}(t)\smash{-}\sqrt{R}\sin\Phi_{1}(t)],\\
\langle\hat{\mathcal{Y}}_{1}(t)\rangle&\smash{=}&|\alpha_{0}(t)|e^{-\mu(t)}
[\sqrt{T}\sin\varphi_{2}(t)\smash{+}\sqrt{R}\cos\Phi_{1}(t)],
\end{eqnarray}
\end{mathletters}
where $\Phi_{1}(t)=\psi(t)+\varphi_{1}(t)$, $\psi(t)=2\gamma\bar{n}_{1}(t)$,
$\mu(t)=\gamma\psi(t)/2$ and $\varphi_{1}(t)$ is the initial phase of the pulse in
nonlinear section. For the correlation functions of the quadrature components we have
\begin{mathletters}
\begin{eqnarray}
R_{{\mathcal{X}}_1}(t,t+\tau)=\frac{1}{4}\Bigl\{\delta(\tau)&+&R\psi(t)h(\tau)\sin2\Phi_{1}(t)\nonumber\\
&+&R\psi^{2}(t)g(\tau)\cos^{2}\Phi_{1}(t)\Bigl\},\\
R_{{\mathcal{Y}}_1}(t,t+\tau)=\frac{1}{4}\Bigl\{\delta(\tau)&-&R\psi(t)h(\tau)\sin2\Phi_{1}(t)\nonumber\\
&+&R\psi^{2}(t)g(\tau)\cos^{2}\Phi_{1}(t)\Bigl\}.
\end{eqnarray}
\end{mathletters}
For the spectra of quadrature fluctuations we find
\begin{mathletters}
\begin{eqnarray}
S_{{\mathcal{X}}_{1}}(\Omega,t)=\frac{1}{4}\bigl[1&+&2R\psi(t)L(\Omega)\sin2\Phi_{1}(t)\nonumber\\
&{}&+4R\psi^{2}(t)L^{2}(\Omega)\cos^{2}\Phi_{1}(t)\bigl],\label{su1}\\
S_{{\mathcal{Y}}_{1}}(\Omega,t)=\frac{1}{4}\bigl[1&-&2R\psi(t)L(\Omega)\sin2\Phi_{1}(t)\nonumber\\
&{}&+4R\psi^{2}(t)L^{2}(\Omega)\sin^{2}\Phi_{1}(t)\bigl].\label{su2}
\end{eqnarray}
\end{mathletters}
At optimal phase (\ref{phase}) the spectra (\ref{su1}-\ref{su2}) take the forms
\begin{mathletters}
\begin{eqnarray}
S_{{\mathcal{X}}_{1}}(\Omega_{0},t)&=&\frac{1}{4}\Bigl\{\bigl[\sqrt{1+\psi^{2}(t)L^{2}(\Omega_{0})}+R\psi(t)L(\Omega_{0})\bigl]^{2}\nonumber\\
&{}&-\left(2R-1\right)^2\psi^{2}(t)L^{2}(\Omega_{0})\Bigl\},\\
S_{{\mathcal{Y}}_{1}}(\Omega_{0},t)&=&\frac{1}{4}\Bigl\{\bigl[\sqrt{1+\psi^{2}(t)L^{2}(\Omega_{0})}-R\psi(t)L(\Omega_{0})\bigl]^{2}\nonumber\\
&{}&-\left(2R-1\right)^2\psi^{2}(t)L^{2}(\Omega_{0})\Bigl\}.
\end{eqnarray}
\end{mathletters}
At any frequency $\Omega$ we have
\begin{mathletters}
\begin{eqnarray}
S_{{\mathcal{X}}_{1}}(\Omega,t)\!&=&\!S_{{\mathcal{X}}_{1}}(\Omega_{0},t)+\frac{1}{2}\,R\,\psi(t)[L(\Omega)-L(\Omega_{0})]\nonumber\\
&{}&\times\Bigl\{[L(\Omega)+L(\Omega_{0})]\psi(t)+[1+(L(\Omega)\smash{+}L(\Omega_{0}))\nonumber\\
&{}&~\times
L(\Omega_{0})\psi^{2}(t)][1\smash{+}\psi^{2}(t)L^{2}(\Omega_{0})]^{-1}\Bigl\},\label{sar1a}\\
S_{{\mathcal{Y}}_{1}}(\Omega,t)\!&=&\!S_{{\mathcal{Y}}_{1}}(\Omega_{0},t)+\frac{1}{2}\,R\,\psi(t)[L(\Omega)-L(\Omega_{0})]\nonumber\\
&{}&\times\Bigl\{[L(\Omega)+L(\Omega_{0})]\psi(t)-[1+(L(\Omega)\smash{+}L(\Omega_{0}))\nonumber\\
&{}&~\times
L(\Omega_{0})\psi^{2}(t)][1\smash{+}\psi^{2}(t)L^{2}(\Omega_{0})]^{-1}\Bigl\}.\label{sar1b}
\end{eqnarray}
\end{mathletters}

\begin{figure}[t]
   \begin{center}
       \leavevmode
       \epsfxsize=.48\textwidth
       \epsfysize=.4\textwidth
       \epsffile{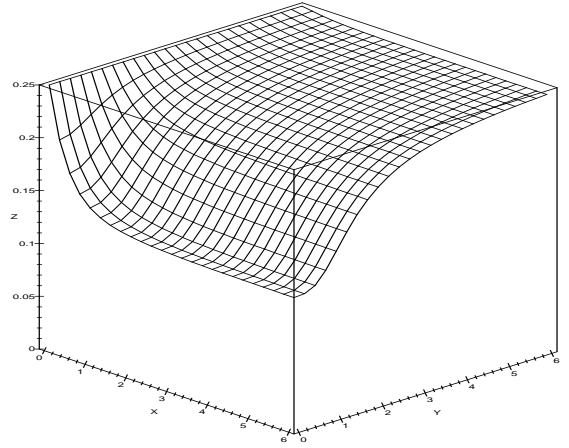}
   \end{center}
   \caption{Dispersion [Z] of the ${\mathcal{Y}}_{1}$- quadrature component of a pulse at time $t\smash{=}0$
as a function of the maximum nonlinear phase $\psi(0)$ [X] and the reduced frequency
$\Omega\smash{=}\omega\tau_r$ [Y], at values of the phase of the pulse which are
optimal for $\Omega_{0}\smash{=}0$ ($R=T=1/2$). \label{fig2}}
\end{figure}

As one can see from (\ref{sar1a}-\ref{sar1b}) the interference of the SPM-USP with
the coherent USP does not influence the quadrature squeezing of the SPM-USP. The
choice of the coefficient of reflection $R$ of the beam-splitter allows us to receive
the spectra of squeezing (see (\ref{sar1a}-\ref{sar1b})) of interest to us. One can
conclude that, the lost of photons from SPM-USP as a result of division in the
beam-splitter is follows by the reduction of squeezing. In fact this lost is
compensated by the coherent USP. The spectra of quantum fluctuations of
${\mathcal{Y}}_{1}$ quadrature at time $t=0$ as a function of maximum nonlinear phase
$\psi(0)=\psi_{0}$ and the reduced frequency $\Omega$ in the case of the $50\%$
beam-splitter ($R=T=1/2$) is displayed in Fig.\ref{fig2}. On Fig.\ref{fig2} one can
see that, in this case the squeezing is reduced at one half.

Also, it is important to remark that, as a result of symmetric reflection the
squeezing from the $X$ quadrature is moved in the $Y$ quadrature. This means that the
ellipse of squeezing is moved in the $XY$ frame with an angle equal to double
reflection angle. {}From theoretical point of view this is expressed by the presence
of the index $i=\exp{(i\pi/2)}$ in the (\ref{transf1}-\ref{transf2}) and
(\ref{transf1a}-\ref{transf2a}). The double angle of reflection can be interpreted as
a geometrical phase. The choice of the position of the beam-splitter relative to
direction of propagation of the SPM-USP allows us to control the position of the
ellipse of squeezing in the  $XY$ frame.
\subsection{The spectra of quantum fluctuations of quadratures at the beam-splitter output \# $2$}
 At the output \# $2$ the quadrature components are defined as
\begin{mathletters}
\begin{eqnarray}
\hat{\mathcal{X}}_{2}(t)&=&[\hat{B}_{2}(t)+\hat{B}^{+}_{2}(t)]/2,\\
\hat{\mathcal{Y}}_{2}(t)&=&[\hat{B}_{2}(t)-\hat{B}^{+}_{2}(t)]/2i.
\end{eqnarray}
\end{mathletters}
For the average values of quadrature components we find
\begin{mathletters}
\begin{eqnarray}
\langle\hat{\mathcal{X}}_{2}(t)\rangle&\smash{=}&|\alpha_{0}(t)|e^{-\mu(t)}
[\sqrt{T}\cos\Phi_{1}(t)\smash{-}\sqrt{R}\sin\varphi_{2}(t)],\\
\langle\hat{\mathcal{Y}}_{2}(t)\rangle&\smash{=}&|\alpha_{0}(t)|e^{-\mu(t)}
[\sqrt{T}\sin\Phi_{1}(t)\smash{+}\sqrt{R}\cos\varphi_{2}(t)].
\end{eqnarray}
\end{mathletters}
Leaving out the preliminary accounts for the correlation functions of the quadrature
components we get
\begin{mathletters}
\begin{eqnarray}
R_{{\mathcal{X}}_2}(t,t+\tau)=\frac{1}{4}\Bigl\{\delta(\tau)&-&T\psi(t)h(\tau)\sin2\Phi_{1}(t)\nonumber\\
&+&T\psi^{2}(t)g(\tau)\cos^{2}\Phi_{1}(t)\Bigl\},\\
R_{{\mathcal{Y}}_2}(t,t+\tau)=\frac{1}{4}\Bigl\{\delta(\tau)&+&T\psi(t)h(\tau)\sin2\Phi_{1}(t)\nonumber\\
&+&T\psi^{2}(t)g(\tau)\cos^{2}\Phi_{1}(t)\Bigl\}.
\end{eqnarray}
\end{mathletters}
In consequence, for the spectra of quantum fluctuations of quadratures we have
\begin{mathletters}
\begin{eqnarray}
S_{{\mathcal{X}}_{2}}(\Omega,t)=\frac{1}{4}\bigl[1&-&2T\psi(t)L(\Omega)\sin2\Phi_{1}(t)\nonumber\\
&{}&+4T\psi^{2}(t)L^{2}(\Omega)\sin^{2}\Phi_{1}(t)\bigl],\label{tur1}\\
S_{{\mathcal{Y}}_{2}}(\Omega,t)=\frac{1}{4}\bigl[1&+&2T\psi(t)L(\Omega)\sin2\Phi_{1}(t)\nonumber\\
&{}&+4T\psi^{2}(t)L^{2}(\Omega)\cos^{2}\Phi_{1}(t)\bigl].\label{tur2}
\end{eqnarray}
\end{mathletters}
The spectra (\ref{tur1}-\ref{tur2}) at optimal phase (\ref{phase}) are:
\begin{mathletters}
\begin{eqnarray}
S_{{\mathcal{X}}_{2}}(\Omega_{0},t)&=&\frac{1}{4}
\Bigl\{\bigl[\sqrt{1+\psi^{2}(t)L^{2}(\Omega_{0})}-T\psi(t)L(\Omega_{0})\bigl]^{2}\nonumber\\
&{}&-(2T-1)^2\psi^{2}(t)L^{2}(\Omega_{0})\Bigl\},\\
S_{{\mathcal{Y}}_{2}}(\Omega_{0},t)&=&\frac{1}{4}
\Bigl\{\bigl[\sqrt{1+\psi^{2}(t)L^{2}(\Omega_{0})}+T\psi(t)L(\Omega_{0})\bigl]^{2}\nonumber\\
&{}&-(2T-1)^2\psi^{2}(t)L^{2}(\Omega_{0})\Bigl\}.
\end{eqnarray}
\end{mathletters}
At any frequency $\Omega$ the spectra (\ref{tur1}-\ref{tur2}) take the forms
\begin{mathletters}
\begin{eqnarray}
S_{{\mathcal{X}}_{2}}(\Omega,t)\!&=&\!S_{{\mathcal{X}}_{2}}(\Omega_{0},t)+\frac{1}{2}\,T\,\psi(t)[L(\Omega)-L(\Omega_{0})]\nonumber\\
&{}&\times\Bigl\{[L(\Omega)+L(\Omega_{0})]\psi(t)-[1+(L(\Omega)\smash{+}L(\Omega_{0}))\nonumber\\
&{}&~\times
L(\Omega_{0})\psi^{2}(t)][1\smash{+}\psi^{2}(t)L^{2}(\Omega_{0})]^{-1}\Bigl\},\label{sar2a}\\
S_{{\mathcal{Y}}_{2}}(\Omega,t)\!&=&\!S_{{\mathcal{Y}}_{2}}(\Omega_{0},t)+\frac{1}{2}\,T\,\psi(t)[L(\Omega)-L(\Omega_{0})]\nonumber\\
&{}&\times\Bigl\{[L(\Omega)+L(\Omega_{0})]\psi(t)+[1+(L(\Omega)\smash{+}L(\Omega_{0}))\nonumber\\
&{}&~\times
L(\Omega_{0})\psi^{2}(t)][1\smash{+}\psi^{2}(t)L^{2}(\Omega_{0})]^{-1}\Bigl\}.\label{sar2b}
\end{eqnarray}
\end{mathletters}

As already have been remarked in the previous analyse, the spectra of quadrature
fluctuations can be controlled by the choice of the coefficients of the beam-splitter
(see (\ref{sar1a}-\ref{sar1b})). It is interesting to remark that, in case of the
symmetric $50\%$ beam-splitter, the squeezing of quadrature fluctuations at output \#
$2$ is present only in the ${\mathcal{X}}_{2}$ quadrature and it is reduced at one
half (see (\ref{sar2a}-\ref{sar2b})). Since the geometrical phase is equal to $0$ in
the analysed case (the refracted part of the SPM-USP does not change the direction of
propagation in comparison with the initial SPM-USP) the ellipse of squeezing does not
change this initial position in the $XY$ frame and has the big axis parallel to $Y$
axis.

\section{The spectra of fluctuations of photon number and the statistics}
We introduce the correlation function of photon number at the output number $j$
($j=\overline{1,2}$) in the following symmetric form
\begin{eqnarray}\label{phot1}
R_{N,j}(t_{1},t_{2})&=&\frac{1}{2}\Bigl[\langle\hat{N}_{j}(t_{1})\hat{N}_{j}(t_{2})\rangle+
\langle\hat{N}_{j}(t_{2})\hat{N}_{j}(t_{1})\rangle\nonumber\\
&{}&-2\langle\hat{N}_{j}(t_{1})\rangle\langle\hat{N}_{j}(t_{2})\rangle\Bigl].
\end{eqnarray}
To simplify the accounts we consider that the initial pulses have the identical
average values of the photon number ($\bar{n}_{1}(t)=\bar{n}_{2}(t)=\bar{n}_{0}(t)$).
Using the algebra of time-dependent Bose-operators (\ref{algebra}-\ref{teor}) for the
correlation functions we get
\begin{mathletters}
\begin{eqnarray}
R_{N,1}(t,t+\tau)\!&=&\!\bar{n}_{0}(t)\Bigl[\delta(\tau)
-2R\sqrt{RT}\psi(t)h(\tau)\cos{\widetilde{\Phi}(t)}\nonumber\\
&{}&+RT\psi(t)h(\tau)\sin{2\widetilde{\Phi}(t)}\nonumber\\
&{}&~+RT\psi^{2}(t)g(\tau)\cos^{2}{\widetilde{\Phi}(t)}\Bigl],\label{kisa1}\\
R_{N,2}(t,t+\tau)\!&=&\!\bar{n}_{0}(t)\Bigl[\delta(\tau)
+2T\sqrt{RT}\psi(t)h(\tau)\cos{\widetilde{\Phi}(t)}\nonumber\\
&{}&+RT\psi(t)h(\tau)\sin{2\widetilde{\Phi}(t)}\nonumber\\
&{}&~+RT\psi^{2}(t)g(\tau)\cos^{2}{\widetilde{\Phi}(t)}\Bigl].\label{kisa2}
\end{eqnarray}
\end{mathletters}
where in (\ref{kisa1}-\ref{kisa2}) is denoted $\tau=t_{1}\smash{-}t$ (see
(\ref{king1}-\ref{king2})), $\widetilde{\Phi}(t)=\psi(t)+\Delta\varphi(t)$,
$\Delta\varphi(t)=\varphi_{1}(t)-\varphi_{2}(t)$. In the case of the symmetrical
$50\%$ beam-splitter, the spectra of quantum fluctuations of the photon number at the
measured time ${\mathcal{T}}$ have the forms
\begin{mathletters}
\begin{eqnarray}
S_{{\mathcal{T}},1}(\Omega,t)\!&=&\!\bar{{\mathcal{N}}}+\bar{n}_{0}(t)\psi(t)L(\Omega)
\left(\frac{\mathcal{T}}{\tau_{p}}\right)
\biggl[-\cos{\widetilde{\Phi}(t)}\nonumber\\
&{}&+\frac{1}{2}\sin{2\widetilde{\Phi}(t)}
+\psi(t)L(\Omega)\cos^{2}{\widetilde{\Phi}(t)}\biggl],\label{spersi1}\\
S_{{\mathcal{T}},2}(\Omega,t)\!&=&\!\bar{{\mathcal{N}}}+\bar{n}_{0}(t)\psi(t)L(\Omega)
\left(\frac{\mathcal{T}}{\tau_{p}}\right)
\biggl[\cos{\widetilde{\Phi}(t)}\nonumber\\
&{}&+\frac{1}{2}\sin{2\widetilde{\Phi}(t)}
+\psi(t)L(\Omega)\cos^{2}{\widetilde{\Phi}(t)}\biggl],\label{spersi2}
\end{eqnarray}
\end{mathletters}
where $\bar{{\mathcal{N}}}=\int_{-\infty}^{\infty}\bar{n}_{0}(t)dt$ is the total
photon number of the initial coherent USP. An extended Mandel parameter at all
frequencies describing the photon statistics at measurement time ${\mathcal{T}}$ can
be introduced as
\begin{mathletters}
\begin{eqnarray}
Q_{{\mathcal{T}},1}(\Omega,t)&=&\varepsilon_{{\mathcal{T}},1}(\Omega,t)\bigl/\bar{n}_{0}(t)\nonumber\\
&=&\psi(t)L(\Omega)\left(\frac{\mathcal{T}}{\tau_{p}}\right)\biggl[-\cos{\widetilde{\Phi}(t)}
\smash{+}\frac{1}{2}\sin{2\widetilde{\Phi}(t)}\nonumber\\
&{}&~+\psi(t)L(\Omega)\cos^{2}{\widetilde{\Phi}(t)}\biggl],\label{speranta1}\\
Q_{{\mathcal{T}},2}(\Omega,t)&=&\varepsilon_{{\mathcal{T}},2}(\Omega,t)\bigl/\bar{n}_{0}(t)\nonumber\\
&=&\psi(t)L(\Omega)\left(\frac{\mathcal{T}}{\tau_{p}}\right)\biggl[\cos{\widetilde{\Phi}(t)}
\smash{+}\frac{1}{2}\sin{2\widetilde{\Phi}(t)}\nonumber\\
&{}&~+\psi(t)L(\Omega)\cos^{2}{\widetilde{\Phi}(t)}\biggl],\label{speranta2}
\end{eqnarray}
\end{mathletters}
where in (\ref{speranta1}-\ref{speranta2}) we denoted
$\varepsilon_{{\mathcal{T}},j}(\Omega,t)\smash{=}S_{{\mathcal{T}},j}(\Omega,t)\smash{-}\bar{{\mathcal{N}}}$.

\begin{figure}[t]
   \begin{center}
       \leavevmode
       \epsfxsize=.48\textwidth
       \epsfysize=.4\textwidth
       \epsffile{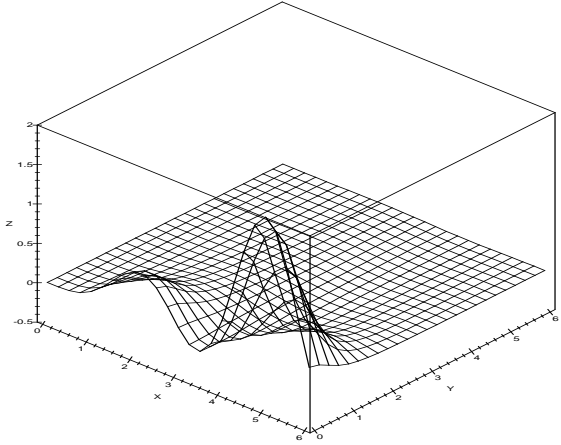}
   \end{center}
   \caption{The dependence of the extended Mandel parameter $Q_{{\mathcal{T}},1}(\Omega,0)$ [Z] on the
maximum nonlinear phase $\psi_{0}$ [X] and the reduced frequency
$\Omega\smash{=}\omega\tau_r$ [Y] at ${\mathcal{T}}/\tau_{p}=1/10$ and
$\triangle\varphi(t)=\pi/2$.\label{mand}}
\end{figure}

The statistics of photon number can be controlled by the choice of the nonlinear
phase addition $\psi(t)$ and linear phase shift $\triangle\varphi(t)$. The dependence
of the extended Mandel parameter $Q_{{\mathcal{T}},1}(\Omega,t)$ at time $t=0$,
${\mathcal{T}}/\tau_{p}=1/10$ and $\triangle\varphi(t)\smash{=}\pi/2$ on $\psi_{0}$
and $\Omega$ is displayed in Fig.\ref{mand}.
As follows from Fig.\ref{mand}, for different values of the nonlinear phase addition,
the statistics of photon number can be as super-Poissonian as sub-Poissonian. The
maximum increase or decrease of photon number fluctuations take place around the
central frequency $\Omega\smash{=}0$. {}From Fig.\ref{mand} it follows that in case
$\triangle\varphi(t)=\pi/2$ the first suppression of the photon number fluctuations
takes place around $\psi(t)=3.5$ and it can be observed at all frequencies from $0$
to $\omega\approx2/\tau_{r}$.

It is important to mention that, if the radiation is monochromatic, the formulas
(\ref{speranta1}-\ref{speranta2}) remain valid. In this case, there are no second and
third terms in (\ref{speranta1}-\ref{speranta2}) and the reduced frequency
$\Omega\smash{=}0$, as since all the photons have the same frequency. The presence of
these terms is connected with the correlation between modes of SPM-USP (see
\cite{POP00b}) and in consequence, the quantum fluctuations of the photon number for
the pulse field will be ``smoothed-out" at the frequency $\Omega\smash{=}0$ in
comparison with the monochromatic radiation.

\section{The modulation of the total photon number}
Let us introduce the photon number operator at the measurement time ${\mathcal{T}}$
\cite{POP00c}
\begin{equation}\label{en}
\hat{{\mathcal{N}}}_{{\mathcal{T}},j}(t)=\int_{t-{\mathcal{T}}/2}^{t+{\mathcal{T}}/2}
\hat{N}_{j}(t^{'})dt^{'},
\end{equation}
where
\begin{equation}
\hat{N}_{j}(t)=\hat{B}^{+}_{j}(t)\hat{B}_{j}(t).
\end{equation}
To simplify the accounts, in following we consider that
$\bar{n}_{1}(t)=\bar{n}_{2}(t)=\bar{n}_{0}(t)$ and then we get
\begin{mathletters}
\begin{eqnarray}
\langle\hat{\mathcal{N}}_{{\mathcal{T}},1}(t)\rangle\!&=&\!{\mathcal{T}}\bar{n}_{0}(t)
\left\{1\smash{-}2\sqrt{RT}\,\sin{\left[\psi(t)\smash{+}\triangle\varphi(t)\right]}\right\},\label{wawwaw1}\\
\langle\hat{\mathcal{N}}_{{\mathcal{T}},2}(t)\rangle\!&=&\!{\mathcal{T}}\bar{n}_{0}(t)
\left\{1\smash{+}2\sqrt{RT}\,\sin{\left[\psi(t)\smash{+}\triangle\varphi(t)\right]}\right\}.\label{wawwaw2}
\end{eqnarray}
\end{mathletters}
where $\bar{n}_{0}(t)=\bar{n}_{0}\rho(t)$ and $\rho(t)$ is the envelope of the pulse.
As can be seen from (\ref{wawwaw1}-\ref{wawwaw2}), for different values of the linear
phase shift $\triangle\varphi(t)$ of the initial coherent light pulses, all photons
can be distributed only in one of the outputs of the beam-splitter. This property is
specific for the quantum interference only and has no classical analogue.

Let us introduce the total number operator at the beam-splitter output \# $1$ as
\begin{equation}\label{enn}
\hat{{\mathcal{N}}}_{1}=\int_{-\infty}^{\infty}\hat{N}_{1}(t)dt,
\end{equation}
and for its average value we obtain
\begin{equation}\label{medium}
\bar{{\mathcal{N}}}_{1}=\bar{{\mathcal{N}}}\left[1-2\sqrt{RT}\,
\frac{\sin(\psi_{0}/4)}{\psi_{0}/4}\,\sin{\left(\frac{\psi_{0}}{4}+\triangle\varphi\right)}\right].
\end{equation}
To get (\ref{medium}) we considered that the initial pulses are quite Gaussian, their
envelopes having the form
$\rho(t)\approx\exp{\left\{-t^{2}\bigl/2\tau^{2}_{p}\right\}}/\sqrt{2}$, and that
$\triangle\varphi(t)\smash{=}\triangle\varphi$ is a constant.

\begin{figure}[t]
   \begin{center}
       \leavevmode
       \epsfxsize=.48\textwidth
       \epsfysize=.35\textwidth
       \epsffile{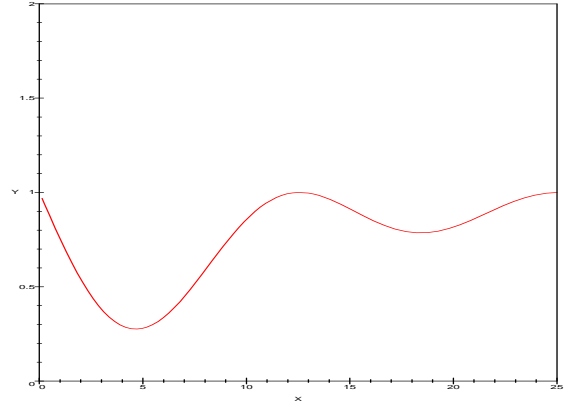}
   \end{center}
   \caption{The dependence of
 $\bar{\mathcal{N}}_{1}/\bar{{\mathcal{N}}}$ [Y] on the maximum
 nonlinear phase $\psi_{0}$ [X] for $\triangle\varphi=0$.
 \label{si}}
\end{figure}

The dependence of $\bar{\mathcal{N}}_{1}/\bar{{\mathcal{N}}}$ at maximum nonlinear
phase addition $\psi_{0}$ for 50\% beam-splitter and $\triangle\varphi\smash{=}0$ is
displayed in Fig.\ref{si}.
As follows from (\ref{medium}) the SPM process leads to the additional modulation of
the total photon number. In consequence, the choice of the maximum nonlinear phase
addition $\psi_{0}$ and the initial phase shift $\triangle\varphi$ allows us to
control the total photon number at the outputs of the beam-splitter.

\section{Discussions and conclusions}
The analyse of the interference process of the SPM-USP with coherent USP, based on
the algebra of time-dependent Bose-operators, allows us to understand in which way
the interference leads to the sub- and super-Poissonian photon statistics formation.
Analysing the spectra of quantum fluctuations of quadrature components was concluded
that, the position of the optical beam-splitter plays an important role in the
quadrature squeezing observation. The initial phase, chosen optimal for a determined
frequency plays a role of the phase of reference. Its choice means that, the $XY$
frame is placed so as the observed squeezing of quadrature $X$ is maximum. As a
result of reflection of the SPM-USP at the beam-splitter, the ellipse of squeezing
will move itself in the $XY$ frame with an angle equal to the geometrical phase. The
geometrical phase can take the values for which the squeezing cannot be observed in
$X$ or $Y$ quadratures. Let us take into consideration the case in which we get the
initial phase of the light pulse in the nonlinear section arbitrary. In this case the
ellipse of squeezing is located arbitrary in the $XY$ frame and it is possible do not
observe the squeezing of quadrature fluctuations in the measurements. In case we
chose the initial phase optimal, we locate the ellipse of squeezing in the $XY$ frame
so as the small axis of the ellipse will lie along the $X$ axis. In consequence, the
observed squeezing of $X$ quadrature fluctuations is maximum. If we do not know the
position of the ellipse in the $XY$ frame (taking the initial phase arbitrary), then
rotating the beam-splitter we can orientate the $XY$ frame so as the ellipse of
squeezing can be displayed with small axis along $X$ axis and the squeezing of the
$X$ quadrature fluctuations can be observed.

Up to now, it is considered that, after reflection from the beam-splitter, the
squeezing of the quadrature fluctuations can be affected and complete disappeared as
a result of the vacuum fluctuations participation at the other input of the
beam-splitter. This interpretation is not quite correct. It is important to mention
that, the presence of vacuum fluctuations was already taken into account when the SPM
process in the nonlinear inertial medium was analysed from quantum point of view
\cite{POP00b}.

The presented in the recent work analyse gives the indication about the use of the
nonclassical light pulses in gravitational wave detection. There are two points of
interest. One is represented by the choice of the initial phase of a pulse at the
input in nonlinear inertial medium which will determine the position of the ellipse
of squeezing of quadrature fluctuations in the $XY$ frame. This optimal phase must be
interpreted as the phase of reference. Other point is represented by the reflection
of the SPM-USP from the surfaces, as since the addition of the geometrical phase will
rotate the ellipse in the $XY$ frame consequently. These points must be taken into
account when the experiments for gravitational wave detection are implemented using
the USP in nonclassical state.

The extended Mandel parameter is introduced and the photon statistics is scanned at
all frequencies. It is shown that, the statistics of the photon number can be
controlled by choice of the nonlinear phase addition and the linear phase shift of
the initial coherent light pulses. The choice of the linear phase shift represents an
effective method of control of the sub- or super-Poissonian statistics formation and
of the modulation of the total photon number.

It is interesting to mention that, the analyse of the interference process between
two SPM-USPs can leads to the determination of the evolution of the linear parameters
of the ellipse of squeezing as a functions of the nonlinear phase addition. This
analyse will be the subject of another publication.\\

{\sf\Blue The author is grateful to S.~Codoban (JINR, Dubna) for useful discussions
and rendered help. \Black}

\end{document}